\documentclass[aps,pre,twocolumn,longbibliography,superscriptaddress]{revtex4-1}
\usepackage[latin1]{inputenc}
\usepackage[english]{babel}
\usepackage{bm,graphicx,graphics,amsmath,amssymb,epsfig,color}
\usepackage{dcolumn}

\newcommand{\mA}{\mathcal{A}}

\newcommand{\mD}{\mathcal{D}}

\newcommand{\mH}{\mathcal{H}}

\newcommand{\mL}{\mathcal{L}}
\newcommand{\mN}{\mathcal{N}}
\newcommand{\mZ}{\mathcal{Z}}

\newcommand{\be}{\begin{equation}}
\newcommand{\ee}{\end{equation}}
\newcommand{\bea}{\begin{eqnarray}}
\newcommand{\eea}{\end{eqnarray}}
\newcommand{\bse}{\begin{subequations}}
\newcommand{\ese}{\end{subequations}}
\newcommand{\comment}[1]{}
\newcommand{\gcol}[1]{{\color{black} #1}}
\newcommand{\ggcol}[1]{{\color{black} #1}}
\newcommand{\rcol}[1]{{\color{black} #1}}

\begin{document}

\title{Symplectic quantization I:\\ dynamics of
  quantum fluctuations in a relativistic field theory}

\author{Giacomo Gradenigo}
\affiliation{Gran Sasso Science Institute, Viale F. Crispi 7, 67100
  L'Aquila, Italy} 
\email{giacomo.gradenigo@gssi.it}

\author{Roberto Livi} \affiliation{Dipartimento di Fisica e Astronomia
  and CSDC, Universit\`a di Firenze, via G. Sansone 1 I-50019, Sesto
  Fiorentino, Italy} \affiliation{Istituto Nazionale di Fisica
  Nucleare, Sezione di Firenze, via G. Sansone 1 I-50019, Sesto
  Fiorentino, Italy} \affiliation{Consiglio Nazionale delle Ricerche,
  Istituto dei Sistemi Complessi, via Madonna del Piano 10, I-50019
  Sesto Fiorentino, Italy}
\email{livi@fi.infn.it}

%%%%%%%%%%%%%%%%%%%%%%%%%%%%%%%%%%%%%%%%%%%%%%%%%%%%%%%%%%%%%%%%%%%%%%%%%%%%%%%
\begin{abstract}
  We propose here a new symplectic quantization scheme, where quantum
  fluctuations of a scalar field theory stem from two main
  assumptions: relativistic invariance and equiprobability of the
  field configurations with identical value of the action. In this
  approach the fictitious time of stochastic quantization becomes a
  genuine additional time variable, with respect to the coordinate
  time of relativity. This \emph{\ggcol{intrinsic} time} is associated
  to a symplectic evolution in the action space, which allows one to
  investigate not only asymptotic, i.e. equilibrium, properties of the
  theory, but also its non-equilibrium transient evolution.  In this
  paper, which is the first one in a series of two, we introduce a
  formalism which will be applied to general relativity in its
  companion work~\cite{GRADE2021}.
\end{abstract}

%\pacs{
%?
%05.60.Cd,05.70.Ln,05.45.Xt
%}
	
\maketitle
\section{Introduction}
\label{intro}

Beyond the notion that there is no absolute time flowing identically
for observers in different reference frames of the universe, special
relativity has challenged the classical concept of \emph{elementary
  particle}, through the equivalence between mass and energy. In fact,
\emph{fields} rather than particles are the fundamental objects, which
are consistent with a genuine relativistic description. They are
represented as non-smooth functions, characterized by infinitely-many
degrees of freedom, which describe the distribution of matter and/or
radiation in the space-time continuum. In this manuscript we discuss a
new pathway for inferring the origin of quantum fluctuations in a
relativistic field theory, making use of concepts borrowed from
statistical mechanics.  More precisely, we are going to argue that
quantum fluctuations in a relativistic field theory can be derived
from a statistical description relying upon a \emph{microcanonical}
formulation, where the conserved quantity is the action, which is the
only relativistically invariant scalar quantity.  \gcol{Due to some
  additional \emph{``kinetic''} degrees of freedom of the quantum
  field, which is our care to introduce and motivate in the following
  discussion, one is able te set up a formalism where the sequence of
  quantum fluctuations experienced by a relativistic quantum field stems from
  a symplectic dynamics parametrized by an
  additional time variable, rather than from the stochastic process of
  the Parisi-Wu formalism~\cite{PW81}.}\\
There are multiple reasons to support this approach.  The
microcanonical generating function, i.e. the entropy functional,
provides the same predictions of the standard path-integral
formulation (e.g. see~\cite{Ramond,DH88,ZinnJustin}), by which one can
estimate the \emph{equilibrium} correlation functions of the field at
different points in space-time, provided one can establish the formal
equivalence between a microcanonical formulation and a canonical one.
Notice that this ``ensemble equivalence'' is not granted a priori for
any relativistic field theory, in analogy with what happens in models
of classical statistical mechanics: for instance, violations of this
equivalence may emerge as a consequence of localization/condensation
phenomena, induced by some peculiar form \rcol{or symmetries} of the
interaction potential~\cite{GILM21}. On the other hand, a
microcanonical formulation which keeps the total action constant has
no conceptual and physical contraindications for being adopted,
independently of the interaction potential, as long as we make
reference to an isolated system, as, for instance, it is the case of a closed universe.
Moreover, in the natural Minkowski metric the action functional is not
positive definite and this prevents the possibility of establishing a
formal correspondence between the microcanonical entropy and the
corresponding canonical potential (the equivalent of the free-energy)
by a Laplace-transform. Conversely, one can always define a
Fourier-transform of the entropy functional, which straightforwardly
yields the standard Feynman path-integral formulation, once the
Fourier dual variable of the action is attributed the value
${\hbar}^{-1}$. \gcol{It is for this necessity to use the Fourier
  integral transform rather than the Laplace one that, at variance
  with ordinary statistical mechanics, the building blocks of quantum
  field theory are probability amplitudes rather than probabilities.}

The main conceptual implication of our microcanonical approach is that
the action can be used as the generator of a symplectic dynamics,
associated to an~\emph{\ggcol{intrinsic} time} $\tau$, replacing
the~\emph{fictitious time} introduced by stochastic
quantization~\cite{PW81,DH88}.  This \ggcol{intrinsic} time governs
the evolution of quantum fluctuations and it has to be distinguished
from the fourth coordinate of the relativistic space-time.

\ggcol{The main advantage of introducing symplectic dynamics
  in~\emph{intrinsic time} is that this provides a consistent (Lorentz
  invariant) and natural framework to study the non-equilibrium
  relaxational dynamics for a relativistic field quantum
  fluctuations. Symplectic quantization is free from the arbitrariness
  usually hidden in the choice of a noise term, typical of any
  Langevin-type dynamics, an arbitrariness which is particularly
  disturbing when noise is used to model the quantum fluctuations of
  relativistic fields. Moreover, symplectic quantization allows to
  define a functional approach to quantum field theory [see, e.g.,
    Eq.~(\ref{eq:Z-micro}) and Eq.~(\ref{eq:micro-prob})] which is
  well defined and physically motivated irrespectively to the
  appearance of infinities from the path integral.}\\

%% . The first one is that we have at
%% disposal a very natural framework for describing the transient
%% out-of-equilibrium dynamics of a relativistic quantum field, which in
%% the limit $\tau\to\infty$ is expected to provide the same statistical
%% inferences of the stochastic quantization approach by Parisi and Wu
%% ~\cite{PW81}.  The second one is that, being the symplectic evolution
%% constrained to a fixed constant value of the action functional, all
%% the divergences (that typically have to be cured by renormalization
%% techniques) inherent in the path-integral formulation are avoided,
%% since they depend on the action taking arbitrary large values.\\

\ggcol{In what follows we provide a mathematical illustration of the
  program just outlined}. In Section~\ref{sec:1} we go through the
main steps of the standard stochastic quantization, making reference,
for the sake of simplicity, to the basic example of a scalar field,
subject to a self-interaction potential.  In Section \ref{sec:2} we
discuss how a symplectic dynamical formalism can be introduced for the
Euclidean formulation of the Lagrangian density and show that one can
pass to the more natural formulation in terms of the action, expressed
in its natural Minkowski metric. Comments and perspectives are
contained in Sec.~\ref{sec:3}.

\section{Stochastic Quantization}
\label{sec:1}

Since our method is inspired by the stochastic quantization by
Parisi-Wu~\cite{PW81}, we first summarize it \gcol{by
  exploiting the elementary example} of a neutral relativistic
invariant scalar field subject to a self-interaction potential $
V(\phi)$. The Lagrangian density of this model is:

\begin{align}
  & \mathcal{L}(\phi,\partial_\mu\phi)  = \frac{1}{2} \partial^\mu \phi\partial_\mu\phi - V(\phi) \nonumber \\
%  \frac{1}{2} m \phi^2 - \frac{\lambda}{4!}\phi^4 \nonumber \\
  & = \frac{1}{2} \left( \frac{\partial \phi}{\partial x^0} \right)^2 - \frac{1}{2} \sum_{i=1}^3 \left( \frac{\partial \phi}{\partial x^i} \right)^2 -  V(\phi)
%  \frac{1}{2} m \phi^2- \frac{\lambda}{4!}\phi^4
  \, , 
\label{eq:lagrangian}
\end{align}
where $x_0=ct$ denotes the~\gcol{coordinate time}. The textbook case
is the nonlinear Klein-Gordon model, where $V(\phi) = \frac{1}{2} m
\phi^2 + \frac{\lambda}{4!}\phi^4 $. The integral over the space-time
continuum of the Lagrangian density in Eq.~(\ref{eq:lagrangian}) is a
relativistic invariant, the action
\begin{align}
S[\phi] = \int dx^0~d{\bf x}~\mathcal{L}(\phi,\partial_\mu\phi) \, .
\end{align}
Due to the Maupertuis \emph{least action} principle, the
\emph{classical} configurations of the field are those which extremize
the action:
\be
\frac{\delta S}{\delta \phi(x^0,{\bf x})} = 0 \, .
\label{eq:least-action}
\ee

According to the path-integral formalism, the generating function for
all the correlations among quantum fluctuations of the theory is the
partition function
\begin{align}
\mZ = \int \mD\phi~\exp\left\lbrace \frac{i}{\hbar}~S[\phi] \right\rbrace \, .
\end{align}
If one then considers the coordinate time $x^0$ as a variable in the
complex plane and performs a Wick rotation, introducing the new
variable $\tilde{x}^0=e^{i\pi/2} x^0 = ix^0$ the Lagrangian density
changes to
\begin{align}
  \mathcal{L}_E &= - \left[  \frac{1}{2} \left( \frac{\partial \phi}{\partial \tilde{x}^0} \right)^2 + \frac{1}{2} \sum_{\mu=1}^3 \left( \frac{\partial \phi}{\partial x^i} \right)^2 + V(\phi)
%   \frac{1}{2} m \phi^2 + \frac{\lambda}{4!}\phi^4 
   \right] \nonumber\\
  &= - \mathcal{L}^{(+)}(\phi,\partial_\mu\phi) \,,
  \label{eq:rotated-lagrangian}
\end{align}
where we have denoted the term between square brackets on the right
hand side of Eq.~(\ref{eq:rotated-lagrangian}) as
$\mathcal{L}^{(+)}(\phi,\partial_\mu\phi)$, to indicate that it
corresponds to a Lagrangian with the wrong sign of the potential
energy. In fact, if we really regard $\tilde{x}^0$ just as an
additional spatial coordinate,
$\mathcal{L}^{(+)}(\phi,\partial_\mu\phi)$ corresponds to a Hamiltonian
density without the kinetic energy, since all terms with derivatives
formally belong to a \emph{potential} energy term.  After Wick's
rotation the action reads
\begin{align}
S[\phi] = i \int d\tilde{x}^0d{\bf x}~\mathcal{L}^{(+)}(\phi,\partial_\mu\phi),
\end{align}
and the corresponding partition function has the form
\begin{align}
  \mZ &= \int \mD\phi~\exp\left\lbrace - \frac{1}{\hbar}~ \int d^4x~\mathcal{L}^{(+)}(\phi,\partial_\mu\phi)\right\rbrace \nonumber \\
  &= \int \mD\phi~\exp\left\lbrace - \frac{1}{\hbar}~S^{(+)}[\phi] \right\rbrace \, .
  \label{eq:Z-rotated}
\end{align}
From Eq.~(\ref{eq:Z-rotated}) it is clear the formal analogy with a
statistical mechanics model where $S^{(+)}[\phi]$ plays the role of
the Hamiltonian and $\hbar$ the one of temperature. The multipoint
correlation functions read:
\begin{align}
  & \left\langle \phi(x) \phi(y) \ldots \phi(z) \right\rangle = \nonumber \\ 
  & = \frac{1}{\mZ} \int \mD\phi~e^{-(1/\hbar)S^{(+)}[\phi]} \phi(x) \phi(y) \ldots \phi(z)
  \label{eq:eucl-corr} 
  \end{align}
The main argument for establishing a relation between \gcol{the
  Euclidean path-integral of Eqs.~(\ref{eq:Z-rotated})
  and~(\ref{eq:eucl-corr})} and stochastic quantization~\cite{PW81} is
very simple (e.g., see \cite{PW81,RSS,DH88}). In order to sample the
configurations of the field $\phi(x)$ with probability
$\exp\lbrace-(1/\hbar)S^{(+)}[\phi]\rbrace$ one usually sets up a
Monte Carlo simulation exploiting the above Boltzmann weight to build
a Metropolis update rule. On the other hand, this \gcol{fictitious
  stochastic process} is essentially equivalent to define a stochastic
dynamics based on a Langevin equation, where it appears a \emph{new}
time variable $\tau$, which in~\cite{PW81} was attributed the
interpretation of a \emph{fictitious time}. \ggcol{Since $\tau$
  parametrizes the ``internal'' dynamics of quantum fluctuations
  independently to the choice of the reference frame}, here we prefer
to call it \emph{\ggcol{intrinsic} time}. \gcol{Stochastic
  quantization amounts to introduce the following} Langevin dynamics
\begin{align}
\frac{\partial}{\partial \tau}\phi(x,\tau) = - \frac{\delta S^{(+)}}{\delta\phi(x,\tau)} + \eta(x,\tau) \, ,
\label{eq:Lang}
\end{align}
where we have made explicit the dependence of the field also on $\tau$ and  $\eta(x,\tau)$ is a
Gaussian random field, defined by its mean value and by its
correlation function:
\begin{align}
   \langle \eta(x,\tau)\rangle &= 0 \nonumber \\
   \langle \eta(x,\tau) \eta(y,\tau') \rangle &= 2 \hbar~\delta^{(4)}(x-y)~\delta(\tau-\tau') 
\end{align}

\gcol{The stochastic equation (\ref{eq:Lang}) was proposed
  in~\cite{PW81} as an alternative approach for sampling in the limit
  $\tau\rightarrow\infty$ the relativistic quantum fields according to
  their Euclidean equilibrium probability measure. In particular the
  main motivation of the stochastic quantization approach was that,
  when dealing with gauge fields, it does not require any gauge-fixing
  procedure~\cite{PW81}. This is why attempts in the direction of
  quantizing also gravity with this approach~\cite{R86} were made
  shortly after the original proposal of the method for non-abelian
  gauge theories~\cite{PW81}. The generalization of the symplectic
  quantization method presented in this paper to gravity will be the
  main topic of~\cite{GRADE2021}.}  \rcol{The most important
  prescription of stochastic quantization is that multipoint
  correlation functions such as $\left\langle \phi(x) \phi(y) \ldots
  \phi(z) \right\rangle$ can be obtained by averaging over the
  stochastic trajectories generated by the Langevin dynamics
  (\ref{eq:Lang}) in the limit of infinite time,
  $\tau\rightarrow\infty$: in formulae
\begin{align}
\lim_{\tau\rightarrow\infty} \left\langle \phi(x,\tau) \phi(y,\tau)
\ldots \phi(z,\tau) \right\rangle_\mD = \left\langle \phi(x) \phi(y)
\ldots \phi(z) \right\rangle,
\label{eq:multipoint}
\end{align}
where the subscript $\mD$ denotes the averages over stochastic
trajectories.  The task of performing explicit calculations by this
method can be accomplished making use of the
Martin-Siggia-Rose-DeDominicis-Janssen
formalism~\cite{MSR73,J76,DD78,ZinnJustin}: in practice the procedure
amounts to solve the problem of a Fokker-Planck operator which, for
$\tau=\infty$, projects onto the stationary Euclidean weight in
Eq.~(\ref{eq:Z-rotated}) (for details see~\cite{PW81}). }

\section{Symplectic quantization}
\label{sec:2}

In order to proceed in the direction of an alternative approach to
stochastic quantization, let us observe that the stochastic formalism
of Eq.~(\ref{eq:Lang}) allows one to introduce a key concept: the
\emph{non-equilibrium relaxation dynamics} for a relativistic quantum
field, which asymptotically approaches an equilibrium value of the
field correlation functions. The dynamical process of quantum
fluctuations which brings the relativistic field across its
\emph{phase-space} with probability $\exp(-S^{(+)}[\phi]/\hbar)$ is
clearly parametrized only by the \ggcol{intrinsic} time $\tau$, while
$x^0=ct$ can be formally considered as a standard coordinate.

In order to illustrate what we mean by \emph{non-equilibrium} in the
context of quantum field theory let us recall first the physical
meaning of the \emph{equilibrium} propagator:
\begin{align}
G(y^0,{\bf y}|x^0,{\bf x}) = \langle \phi(x^0,{\bf x}) \phi(y^0,{\bf y}) \rangle. 
\label{eq:prop-eq}
\end{align}
The expression in Eq.~(\ref{eq:prop-eq}) represents a measurement of
\emph{how much} a certain value of the field at coordinate time $x^0$
and position ${\bf x}$ is correlated to a value of the same field at
another coordinate time $y^0>x^0$ and position ${\bf y}$. This
correlation in space-time is usually evaluated assuming the
\emph{equilibrium} distribution $\exp(-S^{(+)}[\phi]/\hbar)$. But
there might be special situations, e.g. relaxation from very unlikely
configurations following the dynamics of Eq.~(\ref{eq:Lang}), where,
by repeating many times the measurements at time $x^0$ and position ${\bf
  x}$ and comparing them with the measurements of the same field at the
following time $y^0$ and position ${\bf y}$ (it could have been even
${\bf y}={\bf x}$), one may possibly find results not in agreement
with the probability measure $\exp(-S^{(+)}[\phi]/\hbar)$. Assuming
that the theory is correct, a possible explanation of such a behavior
is that the field has \gcol{not yet reached \emph{typical}
  configurations with respect to the euclidean path integral
  measure}. Of course this situation, where the field has not yet
reached equilibrium, must be parametrized by another time variable,
different from the coordinate time that we use to locate events in the
space-time continuum. Such a variable cannot be anything else than the
\ggcol{intrinsic} time $\tau$ of stochastic quantization. \gcol{Such a
  \emph{non-equilibrium} regime would be described by the full
  time-dependent solution (usually out of reach) of the Fokker-Planck
  equation associated to Eq.~(\ref{eq:Lang}). From the knowledge of
  this solution we would have a knowledge of the \emph{time-dependent}
  propagator:}
\begin{align}
  G(y^0,{\bf y};\tau~|~x^0,{\bf x};\tau) = \langle \phi(y^0,{\bf y};\tau)~\phi(x^0,{\bf x};\tau) \rangle.
\end{align}

Coming back to what described in the previous section, one can
interpret $\mL^{(+)}(\phi,\partial_\mu\phi)$ in
Eq.~(\ref{eq:rotated-lagrangian}) as the potential part of the
following Hamiltonian density (see also~\cite{RSS85}):

\begin{align}
  \mH_E(\phi,\pi) &= T -\mL_E \nonumber \\
  & = \frac{1}{2} \frac{1}{c_m^2}\left(\frac{\partial \phi}{\partial \tau}\right)^{2} + \frac{1}{2} \sum_{\mu=0}^3 \left( \frac{\partial \phi}{\partial x^\mu} \right)^2 + V(\phi)
%   \frac{1}{2} m \phi^2 + \frac{\lambda}{4!}\phi^4 
   \nonumber \\
  &= \frac{1}{2} \pi^2(x) + \mL^{(+)}(\phi,\partial_\mu\phi),
  \label{eq:Hamiltonian}
\end{align}
where the momentum
\begin{align}
  \pi(x,\tau)=\frac{1}{c_m}\frac{\partial \phi}{\partial \tau}
\end{align}
is canonically conjugated to $\phi(x,\tau)$ and the subscript $E$ in
$\mH_E(\phi,\pi)$ denotes the \emph{Euclidean} version of our effective
Hamiltonian.  The constant $c_m$, which has to be made explicit for
dimensional reasons, is a velocity. In this framework, it can be
assumed as a free parameter of the theory. On the other hand, lacking
in a relativistic invariant scalar field theory physical parameters
other than $c$ and $\hbar$, there is \emph{a priori} no reason for
preventing the equivalence $c_m=c$, which therefore we will consider
in what follows.\\

We can thus introduce the Euclidean action associated to the
\emph{effective} Hamiltonian density $\mH_E(\phi)$:
\begin{align}
  \mA_E(\phi,\pi) &= \int d^4x~\mH_E(\phi,\pi) \nonumber \\
  &= S^{(+)}[\phi] + \frac{c^2}{2} \int d^4x~\pi^2(x,\tau)
  \label{eq:euclidean-Action}
\end{align}
In analogy with a classical Hamiltonian formulation, one can define the symplectic dynamics
generated by the action $A_E$ as follows
\begin{align}
  \frac{\partial \phi(x,\tau)}{\partial \tau} &= \frac{\delta \mA_E}{\delta \pi(x,\tau)} = \pi(x,\tau)\nonumber \\
  \frac{\partial \pi(x,\tau)}{\partial \tau} &= - \frac{\delta \mA_E}{\delta \phi(x,\tau)} = - \frac{\delta S^{(+)}}{\delta\phi(x,\tau)}\nonumber \\
  \label{eq:Hamilton-eq}
\end{align}
Thus, it turns out that the dynamics of quantum fluctuations in a
relativistic field theory is ultimately ruled by a symplectic
dynamics. \gcol{Then, on the basis of the same assumption of ``bona
  fide'' ergodicity, which is usually taken for granted in most
  Hamiltonian systems, we can postulate the existence of a
  microcanonical ensemble for the fluctuations of the relativistic
  field $\phi(x)$.}  That is, we assume the following \emph{``microcanonical''} postulate:\\

\emph{``All the configurations of the field $\phi(x)$ and the
  conjugated momentum $\pi(x)$ corresponding to an identical value of
  the action $\mA_E(\phi,\pi)$ are attained with identical
  probability''}.\\

Accordingly, the \emph{equilibrium} field theory can be obtained from a
microcanonical statistical ensemble with \gcol{partition volume}:
\begin{align}
\Omega_E(A) = \int \mD\phi \mD\pi~\delta\left(A - \mA_E(\phi,\pi)\right), 
\label{eq:Z-micro-E}
\end{align}
where $A$ denotes a given value of the action. The entropy, i.e. 
the generating functional of all connected diagrams of the relativistic field theory, 
reads 
\begin{align}
\Sigma_E = k~\log \Omega_E(A).
\end{align}

On the other hand, after having sketched above the construction of a
\emph{relativistic invariant microcanonical approach} to field-theory,
one can realize that Wick's rotation, introduced for obtaining a
statistical measure analogous to the Boltzmann-Gibbs canonical measure,
is unnecessary in this context. Accordingly, nothing prevents us from
making a step backward to the original field theory, equipped with its
natural Minkowski metric, and considering the corresponding
Hamiltonian density

\begin{align}
  & \mH_M(\phi,\pi) = T - \mL \nonumber = \\
  &= \frac{c^2}{2} \pi^{2}(x,\tau) - \frac{1}{2}\left( \frac{\partial \phi}{\partial x^0}
  \right)^2 + \frac{1}{2} \sum_{\mu=1}^3 \left(
  \frac{\partial \phi}{\partial x^i} \right)^2 + V(\phi)
%  \frac{1}{2} m \phi^2  + \frac{\lambda}{4!}\phi^4,
 \label{eq:minkovsi-H}
\end{align}
yielding the Minkowski \emph{generalized} action
\begin{align}
  & \mA_M(\phi,\pi) = \int d^4x~\mH_M(\phi,\pi) \nonumber \\
  & = \frac{c^2}{2} \int d^4x~\pi^2(x,\tau) - S[\phi]
  \label{eq:minkovsi-Action}
\end{align}

which allows us to define the corresponding \gcol{partition volume}
\begin{align}
\Omega_M(A) = \int \mD\phi \mD\pi~\delta\left(A - \mA_M(\phi,\pi) \right),
\label{eq:Z-micro}
\end{align}
with the entropy 
\begin{align}
\label{entropy}
  \Sigma_M = k~\log[\Omega_M(A)].
\end{align}
Analogously, the symplectic dynamics (\ref{eq:Hamilton-eq}) changes to
\begin{align}
  \frac{\partial \phi(x,\tau)}{\partial \tau} &= \frac{\delta \mA_M}{\delta \pi(x,\tau)} = \pi(x,\tau)\nonumber \\
  \frac{\partial \pi(x,\tau)}{\partial \tau} &= - \frac{\delta \mA_M}{\delta \phi(x,\tau)}
  \, . 
  \label{eq:Hamilton-eqM}
\end{align} 

Some important remarks are necessary. First of all, we have to
point out that, at variance with $\mA_E$, the Minkowski action $\mA_M$ is
\emph{not a definite sign quantity}, as one can immediately infer from
(\ref{eq:minkovsi-H}) and (\ref{eq:minkovsi-Action}).  Accordingly, in
principle one is not allowed to follow the formal classical pathway to
recover a canonical partition function $\mZ(\beta)$ by
Laplace-transforming \gcol{$\Omega_M(A)$}:
%%, where $\beta$ is the real Lagrange
%%multiplier dual to the action $A$ satisfying the equation
%%
%% \begin{align}
%% \langle A \rangle_\beta = -\frac{\partial}{\partial\beta} \log\left[ \mathcal{Z}(\beta)  \right].
%% \label{saddle-point}
%% \end{align}
%% %%
%%
\gcol{\begin{align}
  \mathcal{Z}(\beta) = \int_0^\infty~dA~e^{-\beta A}~\Omega_M(A).
  \label{eq:Lap-transf}
\end{align}
The Laplace transform of Eq.~(\ref{eq:Lap-transf}) is well defined,
for either positive or negative values of $\beta$, only when the
variable $A$ has a definite sign, which is not the case for a
relativistic field theory.}
%%
%% Moreover, even if one considers
%% configurations of the field $\phi(x,\tau)$ such that $A_M[\phi] > 0$,
%% a solution of Eq.(\ref{saddle-point}) for a positive real value of
%% $\beta$ may not exist: in classical statistical mechanics this is the
%% typical signature of inequivalence between statistical ensembles.
%%
\gcol{ As a consequence, the only transform that one can consistently
  define in complete generality (i.e., not limited to a specific
  interval of values of $A$) is the Fourier-transform $\mZ(z)$:}
\begin{align}
  \mZ(z) &= \frac{1}{\sqrt{2\pi}}\int_{-\infty}^\infty dA~e^{-i z A}~\Omega_M(A) \nonumber \\
  &= \int \mD\phi \mD\pi~e^{-i z \mA_M(\pi,\phi)}.
  \label{eq:Zz}
\end{align}

It is for this reason that the building blocks of quantum field theory
are probability amplitudes rather than probabilities. Due to the
separable form of $\mA_M(\pi,\phi)$, the integration over momenta can
be straightforwardly performed because of its quadratic dependence on
$\pi(x)$ and the result can be hidden inside the infinite
normalization constant $\mN$, typical of field theories. Now, if we
fix $z=1/\hbar$, we recover the standard form of the Feynman path
integral:
\begin{align}
  \mZ(\hbar^{-1}) &=\int \mD\phi \mD\pi~e^{\frac{i}{\hbar} S[\phi] - \frac{i}{\hbar}\frac{c^2}{2} \int d^4x~\pi^2(x,\tau)} \nonumber \\
  &=  \mN(\hbar) \int \mD\phi~e^{\frac{i}{\hbar} S[\phi]}
  \label{eq:Feynman-last}
\end{align}
%%
%\gcol{
The role played by $\hbar$ is analogous, with respect to the
conservation of the generalized action in
Eq.~(\ref{eq:minkovsi-Action}), to the one played by temperature in
the canonical ensemble of statistical mechanics with respect to the
conservation of energy. Accordingly, the role of the canonical
partition function is played by the Feynman path-integral, while the
role of the microcanonical partition sum is played by the expression
in Eq.~(\ref{eq:Z-micro}). \ggcol{Let us point out that the
  microcanonical partition function in Eq.~(\ref{eq:Z-micro}) is not
  merely an equivalent way to write the Feynman path integral, but it
  has a completely new probabilistic interpretation as a
  field-theoretic functional integral. This is particularly clear if
  we put the theory on a lattice: while the Feynman path integral
  still lacks an interpretation in terms of probability, unless we
  consider the corresponding Euclidean field theory by rotating time
  to the imaginary axis, the microcanonical partition function
  $\Omega_M(A)$ yields immediately a probability for the field
  configurations while keeping $\mA_M[\phi,\pi]$ with its original
  Lorentzian metric:
  \begin{align}
    P_A[\phi,\pi] = \frac{1}{\Omega_M(A)}\delta\left( A - \mA_M[\phi,\pi]\right).
    \label{eq:micro-prob}
  \end{align}

Just to mention one possible application, the symplectic quantization
approach allows one to evaluate by means of lattice field theory techniques the
``equilibrium'' (with respect to the above mentioned symplectic
dynamics) correlations between light-cone events, typically occurring
in the study of parton distribution functions~\cite{Ji13}, which are
typically unaccessible in Euclidean field theory. Moreover, one has
the possibility of studying the~\emph{physical} dynamics of quantum
fluctuations by means of Eq.~(\ref{eq:Hamilton-eqM}), making use, for
instance, of a lattice version of the theory.  Note that an approach
allowing one to study the relaxation of quantum fluctuations in
relativistic fields by means of a deterministic dynamics might be
particularly relevant in the context of inflation
theory~\cite{MMOL94,GLMM94,PRT20,PVWA21}.  In this perspective it is
worth mentioning the recent results on the stochastic quantization
approach to the gravitational field discussed in~\cite{BW20}:
according to these results, the whole era before the end of inflation
is characterized by the impossibility to rotate coordinate time from
immaginary back to the real axis, so that the only variable which
governs the dynamics is the fictitious time of stochastic
quantization, corresponding to our \emph{intrinsic} time.}\\

\ggcol{It thus happens that, while in statistical mechanics ensembles
  at fixed temperature or fixed energy are usually equivalent, this is
  not apparently true for quantum field theories, with action taking
  the place of energy. In quantum field theory it seems that the
  ``\emph{pseudo}-microcanonical fixed-action ensemble'' introduced
  here [Eq.~(\ref{eq:Z-micro})] provides a more robust description of
  the system, in particular it looks like that the path-integral
  formulation is more severly plagued by the appearance of divergences
  with respect to symplectic quantization. The reason is the
  following: while the renormalization procedure in quantum field
  theories results in the existence of an additional action scale to
  be fixed by hand (the invariant energy scale at which the
  predictions of the theory are compared with
  experiments)~\cite{Ramond,ZinnJustin}, symplectic quantization fixes
  such a scale from the very beginning. Moreover, in the context of
  symplectic quatization, functional integrals such as the partition
  function in Eq.~(\ref{eq:Z-micro}) are well defined irrespectively
  to the renormalizability of the theory.} \\

In summary, the path-integral formulation of quantum field theory can
be viewed as a representation dual to symplectic quantization. Within
the path-integral the role of quantum fluctuations is assumed to be
reduced to a natural scale factor, $\hbar$, thus epitomizing the role
of the dynamical degrees of freedom $\pi(x)$ into a normalization
constant. The overall procedure taking from symplectic quantization to
the Feynman path integral looks similar to the ergodicity assumption
in the Boltzmann theory of an ideal gas. But there is an important
difference: in the case of symplectic quantization the dynamical
degrees of freedom $\pi(x)$ really seem to play the role of the hidden
variables invoked by Einstein to allow for a complete causal
description of quantum phenomena.

% But the problem of
%ensemble inequivalence remains for all the other cases, where this correspondence
%cannot be established, as it happens for gravity
%(see~\cite{GRADE2021}).\\

\section{Conclusions and Perspectives}
\label{sec:3}

The approach illustrated in this manuscript aims at recasting the
relations among relativistic field theory, quantum fluctuations and
statistical mechanics. By assuming the \emph{agnostic} point of view
according to which it is not assumed the existence of elementary
degrees of freedom for matter (although our theory might be consistent
with this hypothesis) we have proposed an approach to relativistic
fields based on a statistical ensemble characterized by a fixed value
of the action, rather than of the energy. Our approach was inspired by
Parisi-Wu stochastic quantization~\cite{PW81}, where the
\emph{dynamics} of quantum fluctuations for a relativistic field is
parametrized by a fictitious time.

We have discussed how the Parisi-Wu stochastic quantization procedure
is based on an extension of the statistical mechanics ensemble theory,
through the strict analogy between the canonical partition function
and the path-integral formalism, where the role of the temperature is
taken by Planck's constant and the one of the Hamiltonian by the
action of the field theory.  On the other hand, when dealing with a
relativistic field-theory the original action (the one defined in
Minkowski space) is not positive definite and this implies that it
would be much more natural to construct a statistical mechanics
approach exploiting the analogy with a microcanonical equilibrium
ensemble, where the value of the total action fixes the physical scale
of the system, without allowing this quantity to fluctuate over all
its possible values as it happens in a canonical framework. The most
relevant outcome of our analysis is that, once we assume a
microcanonical description, the fictitious time introduced by Parisi
and Wu~\cite{PW81} \ggcol{can be regarded as} a true physical entity:
it is the parameter which rules the evolution of quantum fluctuations
and \ggcol{allows us to consistently extend to quantum field theory
  concepts as \emph{non-equilibrium dynamics}, \emph{relaxation to
    equilibrium} and \emph{irreversibility}. Since this variable is
  related to the ``internal'' dynamics of quantum fluctuations at each
  point of space-time and its definition comes irrespectively to any
  choice of reference frame we called it \emph{intrinsic time}.} Of
course, \emph{coordinate time} is still there, but, as the name says,
it simply plays the role of a coordinate to locate events.  Clearly,
time intervals along the relaxational dynamics parametrized by the
\ggcol{intrinsic} time \ggcol{has to be measured} in terms of the
coordinate time. In order to dispel any doubt about this point, we
want to stress that symplectic quantization does not imply the
existence of any universal time-scale for dynamics. In other words,
symplectic quantization is not associated to the existence of a
universal clock, which would violate the principles of relativity, for
the simple reason that the characteristic time-scale for the dynamics
of $\phi(x,\tau)$ can fluctuate across different regions of the
space-time continuum. \ggcol{In addition, let us point out that the
  elapsing of intrinsic time can be measured only indirectly, for
  instance looking at the irreversible relaxation from very atypical
  configurations of the fields to typical ones. A problem of this kind
  is the one studied in~\cite{DGLR2021} in the context of inflationary
  cosmology.}

%% , as
%% happen for instance, taking an example from the field of condensed
%% matted, for the dynamical heterogeneities in glass-forming
%% systems~\cite{DFGP02,CCGGGPV10,BBBCS11}.

The notion of \emph{\ggcol{intrinsic} time} that we propose here is
quite close to that of \emph{thermal time} proposed in~\cite{CR94}. We
have simply highlighted that, in the case of a relativistic field, a
definition of time conceptually analogous to that of~\cite{CR94} is
perfectly consistent for a relativistic system without any need to
consider a finite temperature.\\

\ggcol{As a test of the new symplectic quantization formalism ideas,
  we have applied them to the study of the gravitational field quantum
  fluctuations in~\cite{GRADE2021}, a work where the presence of a
  cosmological constant term in Einstein equations is put in relation
  with the existence of \emph{``kinetic''} degrees of freedom (in the
  sense of intrinsic time) for the metric field.}\\ \\

\begin{acknowledgments}
  We warmly thank for useful discussions S. Caracciolo, P. Di Cintio,
  S. Matarrese, A. Riotto and A. Vulpiani. In particular we thank
  S. Matarrese for a careful reading of the
  manuscript. G.G. acknowledges Sapienza university of Rome, Physics
  Department, for kind hospitality during some periods in the
  preparation of this manuscript. R.L. acknowledges partial support
  from project MIUR-PRIN2017 \emph{Coarse-grained description for
    non-equilibrium systems and transport phenomena} (CO-NEST)
  n. 201798CZL.

\end{acknowledgments}  

%\bibliography{biblio}

%\bibliographystyle{apsrev}
\end{document}